\documentclass[12pt]{article}

\usepackage{latexsym,amsmath,amssymb}
\usepackage{graphicx}
\usepackage{psfrag}
\usepackage[square, sort&compress, numbers, comma]{natbib}

\newcounter{subequation}[equation]
\makeatletter

\def\bcite{\@ifnextchar [{\@tempswatrue\@bcitex}{\@tempswafalse\@bcitex[]}}
\def\@bcitex[#1]#2{\if@filesw\immediate\write\@auxout{\string\citation{#2}}\fi
  \let\@bcitea\@empty
  \@bcite{\@for\@bciteb:=#2\do
    {\@bcitea\def\@bcitea{,\penalty\@m\ }%
     \def\@tempa##1##2\@nil{\edef\@bciteb{\if##1\space##2\else##1##2\fi}}%
     \expandafter\@tempa\@bciteb\@nil
     \@ifundefined{b@\@bciteb}{{\reset@font\bf ?}\@warning
       {Citation `\@bciteb' on page \thepage \space undefined}}%
     \hbox{\csname b@\@bciteb\endcsname}}}{#1}}
\def\@bcite#1#2{{#1\if@tempswa , #2\fi}}

\def\thesubequation{\theequation\@alph\c@subequation}
\def\@subeqnnum{{\rm (\thesubequation)}}
\def\slabel#1{\@bsphack\if@filesw {\let\thepage\relax
   \xdef\@gtempa{\write\@auxout{\string
      \newlabel{#1}{{\thesubequation}{\thepage}}}}}\@gtempa
   \if@nobreak \ifvmode\nobreak\fi\fi\fi\@esphack}
\def\subeqnarray{\stepcounter{equation}
\let\@currentlabel=\theequation\global\c@subequation\@ne
\global\@eqnswtrue
\global\@eqcnt\z@\tabskip\@centering\let\\=\@subeqncr
$$\halign to \displaywidth\bgroup\@eqnsel\hskip\@centering
  $\displaystyle\tabskip\z@{##}$&\global\@eqcnt\@ne
  \hskip 2\arraycolsep \hfil${##}$\hfil
  &\global\@eqcnt\tw@ \hskip 2\arraycolsep
  $\displaystyle\tabskip\z@{##}$\hfil
   \tabskip\@centering&\llap{##}\tabskip\z@\cr}
\def\endsubeqnarray{\@@subeqncr\egroup
                     $$\global\@ignoretrue}
\def\@subeqncr{{\ifnum0=`}\fi\@ifstar{\global\@eqpen\@M
    \@ysubeqncr}{\global\@eqpen\interdisplaylinepenalty \@ysubeqncr}}
\def\@ysubeqncr{\@ifnextchar [{\@xsubeqncr}{\@xsubeqncr[\z@]}}
\def\@xsubeqncr[#1]{\ifnum0=`{\fi}\@@subeqncr
   \noalign{\penalty\@eqpen\vskip\jot\vskip #1\relax}}
\def\@@subeqncr{\let\@tempa\relax
    \ifcase\@eqcnt \def\@tempa{& & &}\or \def\@tempa{& &}
      \else \def\@tempa{&}\fi
     \@tempa \if@eqnsw\@subeqnnum\refstepcounter{subequation}\fi
     \global\@eqnswtrue\global\@eqcnt\z@\cr}
\let\@ssubeqncr=\@subeqncr
\@namedef{subeqnarray*}{\def\@subeqncr{\nonumber\@ssubeqncr}\subeqnarray}
\@namedef{endsubeqnarray*}{\global\advance\c@equation\m@ne%
                           \nonumber\endsubeqnarray}

\renewcommand\maketitle{\par
  \begingroup
    \if@twocolumn
      \ifnum \col@number=\@ne
        \@maketitle
      \else
        \twocolumn[\@maketitle]%
      \fi
    \else
      \newpage
      \global\@topnum\z@   
      \@maketitle
    \fi
    \thispagestyle{plain}\@thanks
  \endgroup
  \setcounter{footnote}{0}%
  \global\let\thanks\relax
  \global\let\maketitle\relax
  \global\let\@maketitle\relax
  \global\let\@thanks\@empty
  \global\let\@author\@empty
  \global\let\@date\@empty
  \global\let\@title\@empty
  \global\let\title\relax
  \global\let\author\relax
  \global\let\date\relax
  \global\let\and\relax
}

\makeatother

\DeclareFontFamily{OT1}{rsfs11}{}
\DeclareFontShape{OT1}{rsfs11}{m}{n}{ <-> rsfs11 }{}
\DeclareMathAlphabet{\mathscript}{OT1}{rsfs11}{m}{n}

\numberwithin{equation}{section}

\newcommand{\gtlt}{\mathrel{\raise2.5pt\hbox{\oalign{$\scriptstyle>$\crcr
$\scriptstyle<$}}}}

\newcommand{\pt}{\partial}

\newcommand{\be}{\begin{equation}}
\newcommand{\ee}{\end{equation}}
\newcommand{\nn}{\nonumber}
\newcommand{\bea}{\begin{eqnarray}}
\newcommand{\eea}{\end{eqnarray}}
\newcommand{\bsea}{\begin{subeqnarray}}
\newcommand{\esea}{\end{subeqnarray}}

\renewcommand{\d}{\mathrm{d}}

\def\a{\alpha}

\def\f{\phi}

\def\n{\nu}

\def\t{\tau}

\begin{document}

\begin{titlepage}
\begin{flushright}
{\small
DAMTP-2006-117 \\[-0.5ex]
ITFA-2006-48 }
\end{flushright}
\vspace{.5cm}
\begin{center}
\baselineskip=16pt {\huge   Effective Actions for Heterotic M-Theory
\\ }
\vspace{10mm}
{\large Jean-Luc Lehners$^{\dag}$, Paul McFadden$^{\ddag}$ and Neil Turok$^{\dag}$}
\vspace{15mm}

{\small\it $^\dag$ DAMTP, CMS, Wilberforce Road, CB3 0WA, Cambridge, UK.
\\ \vspace{6pt}
$^\ddag$ ITFA, Valckenierstraat 65, 1018XE Amsterdam, the Netherlands.} \\
\vspace{6mm}
\end{center}

\vspace{8mm} \abstract{ \vspace{0.3cm} We discuss the moduli space
approximation for heterotic M-theory, both for the minimal case of
two boundary branes only,  and when a bulk brane is included. The
resulting effective actions may be used to describe the
cosmological dynamics in the regime where the branes are moving
slowly, away from singularities. We make use of the recently
derived colliding branes solution to determine the global
structure of moduli space, finding a boundary at which the
trajectories undergo a hard wall reflection. This has important
consequences for the allowed moduli space trajectories, and for
the behaviour of cosmological perturbations in the model.}

\vspace{2mm} \vfill \hrule width 2.3cm \vspace{2mm}{\footnotesize
\noindent
\hspace{-9mm}
 E-mail: \texttt{j.lehners@damtp.cam.ac.uk, mcfadden@science.uva.nl, n.g.turok@damtp.cam.ac.uk.} }
\end{titlepage}

\setcounter{page}{2}

\section{Introduction}

The conjectured duality of Ho\v{r}ava and Witten \cite{HW1,HW2}
between eleven-dimensional supergravity compactified on the
orbifold $S^1/\mathbb{Z}_2$ and strongly coupled heterotic string
theory is a key development for fundamental string and M-theory,
for particle phenomenology \cite{Braun:2005nv} and for early
universe cosmology \cite{Khoury:2001wf}. Realistic particle
physics models are obtained by first compactifying six of the ten
spatial dimensions on a Calabi-Yau 3-fold, taken to be smaller
than the orbifold dimension for phenomenological reasons, and then
studying the resulting five-dimensional effective theory
\cite{LOSW1,LOSW2} in which one of the remaining four spatial
dimensions is the orbifold with its boundary branes. In recent
work, Moss has given an improved treatment of the boundary
conditions for bulk fields \cite{Moss1,Moss2}, giving greater
confidence in this general approach.

Cosmological applications of heterotic M-theory emphasise the
importance of letting the branes be fully dynamical
\cite{Khoury:2001wf,Steinhardt:2001st}. If heterotic M-theory is
to describe our universe, however, then the constancy of the
measured coupling constants implies that the volume of the
Calabi-Yau, which is related to the gauge couplings, must have
been stabilised shortly after the big bang.  Similarly, the radion
field measuring the size of the extra dimension must either
decouple efficiently from the matter density, or itself be given a
mass. Nevertheless, it is entirely reasonable to assume that the
branes may have been dynamical over a short period of time just
before and after the big bang.  It is therefore of interest to
derive four-dimensional effective actions for heterotic M-theory
describing the motion of bulk and boundary branes, and many
authors have done so in the past \cite{Palma:2004fh,Gonzalo,
GaugesII,KSHW,Derendinger:2000gy, Brandle:2001ts,Correia:2006pj}.
In particular, we would like to draw attention to \cite{Koyama},
which is closely related to the work reported here.

We will derive the effective action in this paper by developing
the moduli space approximation in light of the recently derived
colliding branes solution of \cite{LMT}. This exact solution of
heterotic M-theory enables us to understand the global structure
of moduli space in more detail. In particular, we find that there
are two different types of boundaries to moduli space: the first
type of boundary, which is well-known, arises for example because
a bulk brane is constrained to remain in between the two orbifold
branes, and therefore the range of the coordinate describing its
location along the orbifold is restricted. However, when a bulk
brane collides with one of the orbifold branes, a small instanton
transition can occur \cite{Ovrut:2000qi}, signalling the appearance of
new, light degrees of freedom which cannot be described in the
original four-dimensional effective theory. In other words, the original
moduli space approximation cannot be trusted at these locations
and an improved treatment is needed.

\begin{figure}[p]
\begin{center}
\hspace{-0.5cm}
\includegraphics[width=17cm]{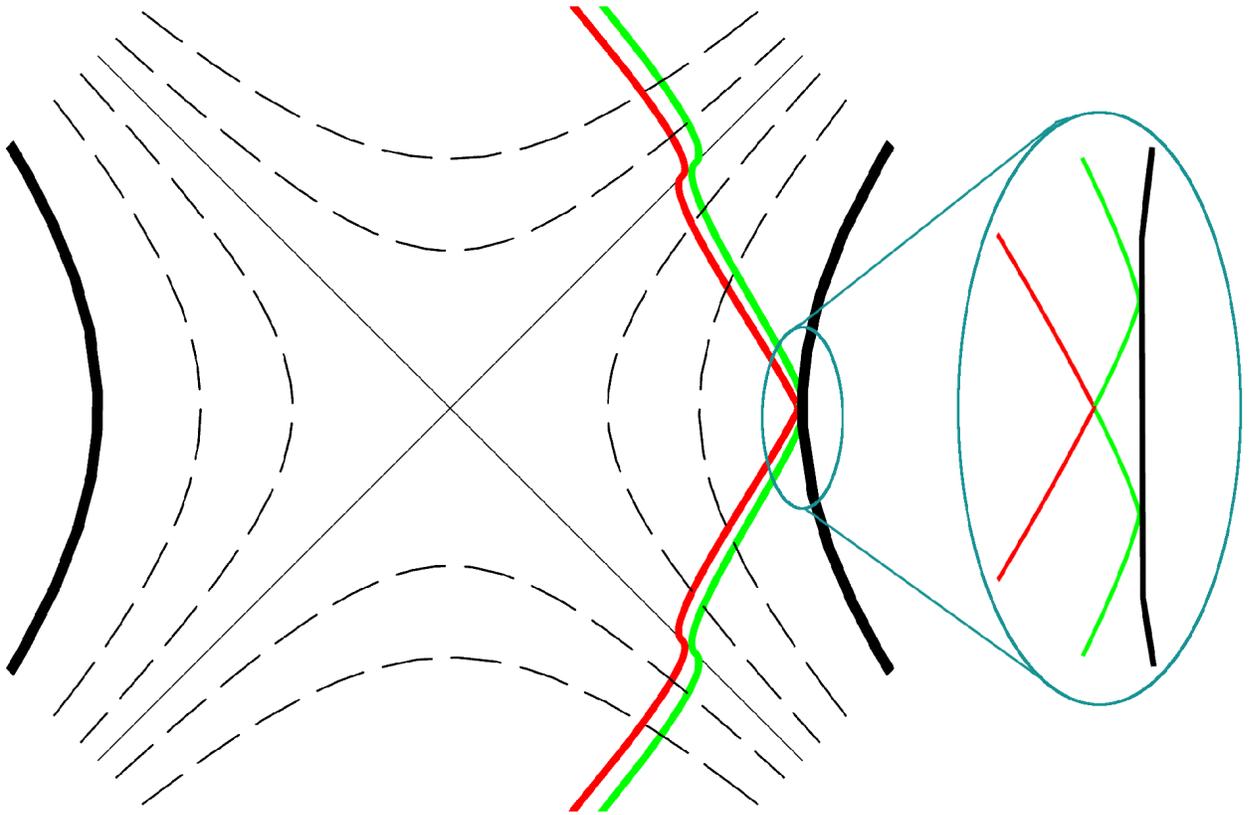}
\vspace{0.5cm}  \caption{ \label{Kruskal} {\small A Kruskal plot of
the colliding branes solution described in \cite{LMT}. In suitable
coordinates, the bulk geometry is static and contains a timelike
naked singularity (denoted by thick black lines) corresponding to
a zero of the bulk warp factor. The dashed lines indicate
representative orbits of the bulk Killing vector field. The
boundary branes then move through this bulk geometry according to
the Israel junction conditions.  The trajectory of the
positive-tension brane is shown in red and that of the
negative-tension brane in green.  The collision of the branes, as
well as the two bounces of the negative-tension brane off the
naked singularity, are shown at a magnified scale in the inset.
The regularisation of these bounces is discussed in detail in
\cite{LMT}. } }
\end{center}
\end{figure}

We will argue, however, that there may also exist boundaries
to moduli space that can best be described as {\it hard} and
repulsive. Such a boundary arises at a zero of the bulk warp factor.
From the higher-dimensional picture, we know that it is always the
negative-tension brane that encounters this singularity, and is in
fact repelled by it (see Figure \ref{Kruskal}).
We describe the boundary as `hard' because in the presence of the
slightest amount of matter on the negative-tension brane, satisfying
the weak energy condition, the singularity is in fact never reached
and the negative-tension brane simply bounces back smoothly. In the
moduli space approximation this corresponds to a simple reflection
of one of the two scalar fields off the boundary. For a small, fixed
density of matter on the negative-tension brane, as we lower the
speed of the incoming negative-tension brane, the bounce becomes
milder and milder. In this situation, we do not expect additional
light degrees of freedom, such as branes wrapping cycles of the
Calabi-Yau, to become light enough to
play a significant role, and we expect the four-dimensional
effective description to remain accurate.

We begin this paper with a brief review of heterotic M-theory, before
proceeding in Section \ref{sectionMSA} to derive the moduli space action for the case in which
there is no bulk brane present.
From this action,
we show how to recover the colliding branes solution recently
discussed in \cite{LMT}. This particular solution highlights
another general feature of the moduli space approximation that shows the
importance of higher-dimensional input: the moduli space
approach leads to an infinite number of possible background solutions. Choosing
a particular one then amounts to imposing boundary conditions that
are obtained from the original higher-dimensional theory.
We will generalise our discussion to
incorporate the presence of a bulk brane in Section \ref{sectionBulk}.  In particular, we exhibit a series of special
cases in which the moduli space action simplifies. 
In these instances, we are able to recover and improve upon some earlier results appearing in the
literature.  We conclude in Section \ref{sectionDiscussion}.

\section{Heterotic M-Theory}

Ho\v{r}ava-Witten theory can be dimensionally reduced on a
Calabi-Yau 3-fold to yield five-dimensional heterotic M-theory
\cite{HW1,HW2,LOSW1,LOSW2}. This gauged supergravity theory can be
consistently truncated to gravity and a scalar $\f$ parameterising
the volume of the Calabi-Yau manifold (namely $V_{\mathrm{CY}} = e^{\f}$).
The action is given by
 \bea
S &=& \int_{5d} \sqrt{-g} \,[R - \frac{1}{2}(\pt \phi)^2 - 6\a^2 e^{-2 \phi}] \nn \\
&& + 12 \a \int_{4d,\,y=-1} \sqrt{-g} e^{- \phi} - 12 \a
\int_{4d,\,y=+1} \sqrt{-g} e^{- \phi}, \label{Action5d} \eea where
$\a$ is related to the number of units of 4-form flux pointing
entirely in the Calabi-Yau directions\footnote{Compared to
\cite{LOSW1, LOSW2}, we have re-scaled $\a$ such that $\a =
\a_{\mathrm{LOSW}}/3\sqrt{2}$.}, and we have placed branes of
opposite tensions at $y=\pm 1$ (where $y$ is the coordinate
transverse to the branes). It is important to realise that when
boundary brane actions are present $\a$ is necessarily non-zero.
As a consequence five-dimensional Minkowski space is not the vacuum
of the theory; rather, the vacuum is given by a domain wall
spacetime of the form \bea
\d s^2 &=& h^{2/5}(y)\,\big[B^2 \,(-\d \t^2 + \d \vec{x}^2) + A^2 \,\d y^2\big], \nn \\
 e^{\f} &=& A\, h^{6/5}(y), \label{domainwall} \nn \\
h(y) &=& 5\a\, y+C,
\eea
where $A$, $B$ and $C$ are arbitrary constants.
The $y$ coordinate is taken to span
the orbifold $S^1/\mathbb{Z}_2$ with fixed points at $y=\pm 1$.
In an `extended' picture of the solution, obtained by $\mathbb{Z}_2$-reflecting the solution across the branes,
there is a downward-pointing kink at $y=-1$ and an upward-pointing kink at $y=+1$.
These ensure the Israel conditions are satisfied, with the negative-tension brane being located at $y=-1$ and
the positive-tension brane at $y=+1$.

\section{The Moduli Space Action}
\label{sectionMSA}

In our previous paper \cite{LMT}, we derived the
higher-dimensional solution for colliding branes in heterotic
M-theory. Here, we wish to describe this solution in terms of the
four-dimensional effective theory. We derive this theory by making
use of the moduli space approximation, originally employed in the
study of the low-energy dynamics of BPS monopoles
\cite{Manton:1981mp,Manton:1985hs}, and subsequently developed
extensively in the mathematical literature \cite{Atiyah:1985dv}.
The main idea is the following. If the set of static solutions of
given topology is parameterised by some continuous parameters,
then these parameters represent flat directions in configuration
space, which cost no potential energy. In contrast, other
directions in configuration space are typically associated with
large mass scales (for example, the massive vector boson mass in a
spontaneously broken gauge theory). Hence the low-energy dynamics
of the system can be well-approximated by considering motion to be
only along the massless directions. It is important to emphasise
that the moduli space approximation is usually {\it not} exact: it
ignores effects due to radiation when two monopoles scatter, for
example. But in many cases, it is found to give the correct
leading order description of the dynamics, in an expansion in the
velocity of the motion along moduli space. In many cases the
system departs from the ``bottom of the potential valley",
described by moduli space, by an amount proportional to the square
of the velocity along moduli space. A simple example is provided
by a a theory with a broken U(1) global symmetry, {\it i.e.} a
complex scalar field $\phi= f e^{i\theta}$ with a Mexican hat
potential. Restricting attention to spatially homogenous
solutions, the low-energy motion consists of the field running
around the potential minimum $f=f_{min}$ with some velocity
$\dot{\theta}$, while the massive field $\delta f =f-f_{min}$,
deviates only modestly from the potential minimum, $(\delta f
/f_{min}) \sim \dot{\theta}^2/m^2$, where $m$ is the mass of
$\delta f$.

Here, following the usual procedure, we shall derive the
moduli space action by plugging the static solutions, with
parameters promoted to time-dependent moduli,
back into the action and integrating
out the spatial dependence. The resulting action contains
only kinetic terms, and from these the metric which
governs geodesic motion on moduli space can be read off.
On general grounds, one expects the low-energy
dynamics, including both mild space and time gradients,
to be described by a 4d effective theory respecting
full spacetime symmetry ({\it i.e.} Lorentz or general coordinate
invariance). Once one has determined the
kinetic terms from the moduli space approach, it is usually
straightforward to identify the corresponding,
fully covariant four-dimensional
spacetime action.

\subsection{The Time-Dependent Moduli}

In the static domain wall solution above, the volume of the
Calabi-Yau manifold and the distance between the boundary branes
are determined in terms of the moduli $A$ and $C$, while the scale
factors on the branes are determined in terms of $B$ and $C$.  The
modulus $C$ additionally determines the height of the harmonic
function $h$ at a given position in $y$. To implement the moduli
space approximation, we simply promote these moduli to arbitrary
functions of the brane conformal time $\t$, yielding the ansatz:
\bea \d s^2 &=&
h^{2/5}(\t,y)\left[B^2(\t)\,(-\d \t^2 + \d \vec{x}^2) + A^2(\t) \d y^2\right], \nn \\
e^{\f} &=& A(\t)\, h^{6/5}(\t,y), \label{TimeMod} \nn \\
h(\t,y) &=& 5\a y +C(\t), \qquad -1 \le y \le +1.
\eea
Let us give a brief justification for this ansatz:
firstly, we note that the ansatz satisfies the $\tau y$ Einstein
equation identically. This is important, since otherwise the $\tau
y$ equation would act as a constraint, see {\it e.g.}
\cite{Gray:2003vw}. Secondly, there is no $g_{\tau y}$ modulus,
since this metric component is odd under the $\mathbb{Z}_2$
symmetry, and therefore has to vanish at the location of the
branes. Any such component which is zero at the location
of the branes, but non-zero in the bulk, is necessarily
massive. In fact, from the work of \cite{Lehners:2005su}, we know
that, apart from the above moduli, all other perturbations have a
positive mass squared.

For completeness, the lift of this ansatz to eleven dimensions is given by
\bea
\d s_{11}^2 &=& e^{-2\f/3} \,\d s_5^2 + e^{\f/3}\, \d s_{CY}^2 \\
&=& h^{-2/5} A^{-2/3} B^{2}\, (-\d\t^2+\d\vec{x}^2) +
h^{-2/5} A^{4/3} \,\d y^2 + h^{2/5} A^{1/3}\,\d s_{\mathrm{CY}}^2,
\eea
where the five-dimensional metric and scalar field are now both part of
the eleven-dimensional metric.
The eleven-dimensional distance between the branes is then
\be
d_{11} = A^{2/3} \,I_{-\frac{1}{5}},
\label{distance11d}
\ee
where we have defined
\bea
I_n &=& \int_{-1}^{1} dy \ h^n = \frac{1}{5\a(n+1)}[(C+5\a)^{(n+1)}-(C-5\a)^{(n+1)}].
\eea
The orbifold-averaged Calabi-Yau volume is given by
\be
\langle e^{\f} \rangle = \frac{1}{2} A I_{\frac{6}{5}}.
\ee

\subsection{The Action} \label{SectionMSA}

Having defined the time-dependent moduli, we would now like to
derive the action summarising their equations of motion. This is
achieved by simply plugging the ansatz (\ref{TimeMod}) into the
original action (\ref{Action5d}), yielding the result (where we
use the notation $\dot{ } \equiv \pt/\pt \t$)
\bea
S_{\mathrm{mod}} &=& -6 \int_{4d}
AB^2I_{\frac{3}{5}} \left[ -\frac{1}{12}\Big(\frac{\dot{A}}{A}\Big)^2
+\Big(\frac{\dot{B}}{B}\Big)^2 +\frac{\dot{A}\dot{B}}{AB}
-\frac{1}{25}\frac{I_{-\frac{7}{5}}}{I_{\frac{3}{5}}}\,\dot{C}^2
+\frac{3}{5}\,\frac{I_{-\frac{2}{5}}\dot{B}\dot{C}}{I_{\frac{3}{5}}B}\right].\qquad
\label{ActionMSA1}
\eea
This action can be greatly simplified by
introducing the field redefinitions
\bea
a^2 &\equiv& A\,B^2\,I_{\frac{3}{5}}, \\
e^{\sqrt{3}\psi} &\equiv& A\,(I_{\frac{3}{5}})^{3/4}, \\
\chi &\equiv& -\frac{1}{20}\int \d C\, (I_{\frac{3}{5}})^{-1}\,
\big[9\,(I_{-\frac{2}{5}})^2+16\,I_{-\frac{7}{5}}I_{\frac{3}{5}}\big]^{1/2}.
\label{DefinitionChi}
\eea
Note that $a$ has the interpretation of
being roughly the four-dimensional scale factor, whereas $\psi$ and
$\chi$ are four-dimensional scalars. The definition
(\ref{DefinitionChi}) can be rewritten as stating that
\be
\d\chi = -\frac{\d C}{2\,(C+5\a)^{1/5}\,(C-5\a)^{1/5}\, I_{\frac{3}{5}}}.
\ee This expression can be integrated to yield \be C = 5\a \left[
\frac{(1+e^{4\chi})^{5/4}
+(1-e^{4\chi})^{5/4}}{(1+e^{4\chi})^{5/4}
-(1-e^{4\chi})^{5/4}}\right]. \label{relCmoduli} \ee In terms of
$a$, $\psi$ and $\chi$ the moduli space action (\ref{ActionMSA1})
then reduces to the remarkably simple form \be S_{\mathrm{mod}} =
6 \int_{4d} [-\dot{a}^2 + a^2 (\dot{\psi}^2 + \dot{\chi}^2)].
\label{ActionMSA2} \ee The minus sign in front of the kinetic term
for $a$ is characteristic of gravity, and in fact this is the
action for gravity with scale factor $a$ and two minimally coupled
scalar fields. There is also a manifest $O(2)$ rotation symmetry
for the scalar fields. The equation of motion for $a$ reads \be
\ddot{a} = - a\,(\dot{\psi}^2 + \dot{\chi}^2), \ee while the
equations of motion for $\psi$ and $\chi$ immediately lead to the
conserved charges $Q_{\psi}$ and $Q_{\chi}$, according to \be
a^2\, \dot{\psi} = Q_{\psi}, \qquad a^2 \,\dot{\chi} = Q_{\chi} .
\ee The solutions to these equations are given by  \bea
a^2 &=& 2\,\sqrt{Q_{\psi}^2+Q_{\chi}^2}\,(\t-\t_a), \\
\psi &=& \frac{Q_{\psi}}{2\,\sqrt{Q_{\psi}^2+Q_{\chi}^2}}\,\ln [\psi_0(\t-\t_a)], \\
\chi &=& \frac{Q_{\chi}}{2\,\sqrt{Q_{\psi}^2+Q_{\chi}^2}}\,\ln [\chi_0(\t-\t_a)],
\eea
where $Q_{\psi}$, $Q_{\chi}$, $\t_a$, $\psi_0$, and
$\chi_0$ are constants of integration.

We can now return to the ansatz (\ref{TimeMod}) and relate
physical quantities in five dimensions to the moduli fields $a$,
$\psi$ and $\chi$: if we denote the distance between the branes by
$d$, and the volume of the Calabi-Yau and the brane scale factors at the locations $y = \pm1$ by
$e^{\f}_{\pm}$ and $b_{\pm}$ respectively,
then we have the relations
\bea
d &=& \frac{1}{3}\,(2\a)^{-1/4}\, e^{-3\chi+\sqrt{3}\psi}\,
[(1+e^{4\chi})^{3/2}-|1-e^{4\chi}|^{3/2}], \label{relModDistance} \\
e^{\f}_{\pm} &=& (2\a)^{3/4} \, e^{\sqrt{3}\psi} \,\left\{
\begin{array}{lll}
                     \left( \cosh 2 \chi\right)^{3/2}  & \\
                     |\sinh 2 \chi|^{3/2}, & \\
                    \end{array} \right.  \\
b_{\pm} &=& (2\a)^{1/8} \,a \,e^{-\sqrt{3}\psi/2} \,\left\{
\begin{array}{lll}
                     \left( \cosh 2 \chi\right)^{1/4}  & \\
                     |\sinh 2 \chi|^{1/4}. & \\
                    \end{array} \right. \label{relModScalefactor}
\eea
These relations are useful in
interpreting particular solutions to the moduli equations of
motion. Note that for $\chi \rightarrow + \infty$, we have \bea
d &\simeq& (2\a)^{-1/4}\, e^{\sqrt{3} \psi - \chi}, \\
e^{\f}_{\pm} &\simeq& (2\a)^{3/4}\, e^{\sqrt{3} \psi + 3 \chi},
\eea
whereas for $\chi \rightarrow - \infty$, we have
\bea
d &\simeq& (2\a)^{-1/4}\, e^{\sqrt{3} \psi + \chi}, \\
e^{\f}_{\pm} &\simeq& (2\a)^{3/4}\, e^{\sqrt{3} \psi - 3\chi}.
\eea Thus, in both limits, $\ln{d}$ and $\f_{\pm}$ are orthogonal
variables. This means that, sufficiently far away from the $\chi =
0$ axis, the fields $\psi$ and $\chi$ are, up to a re-scaling,
simply related to $\ln{d}$ and ${\f_{\pm}}$ by a rotation in field
space. Since the moduli space trajectories in terms of $\psi$ and
$\chi$ are straight lines, far from the $\chi =
0$ axis, the trajectories will also be approximately straight lines in terms
of $\ln{d}$ and $\f_{\pm}$.

\begin{figure}[ht]
\begin{center}
\includegraphics{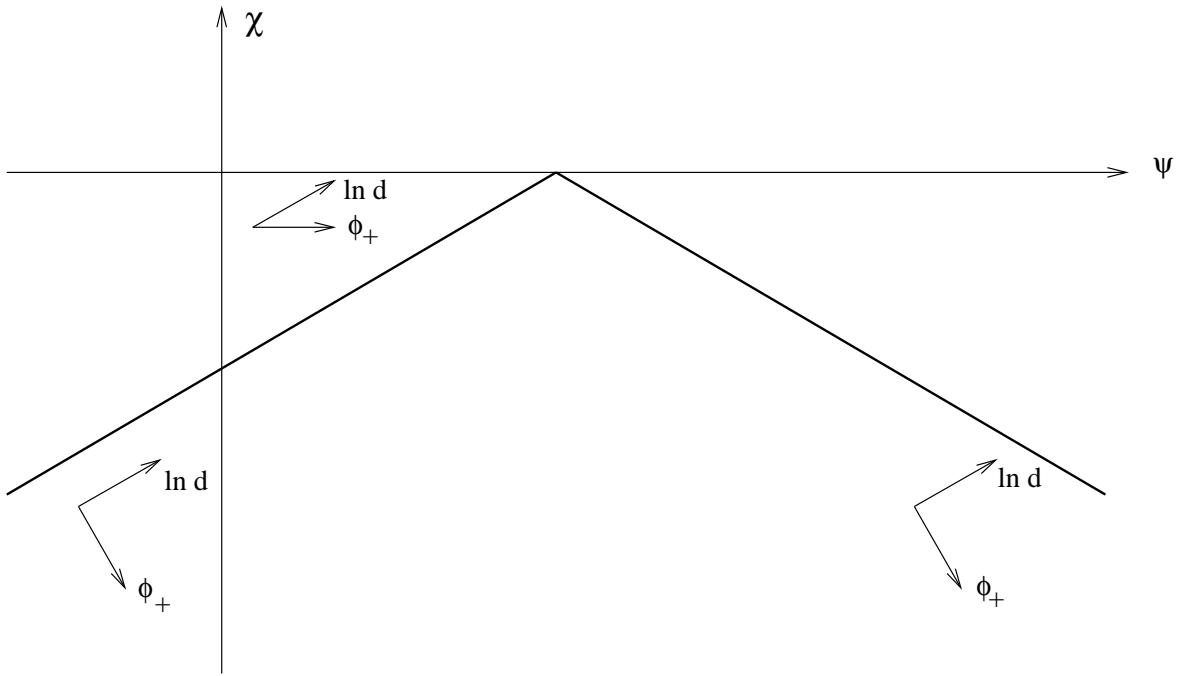}  \caption{\label{fig1} {\small The trajectory of the scaling solution as seen in the $\psi$ - $\chi$ plane.
$\chi = 0$ corresponds to the scale factor on the negative-tension
brane shrinking to zero, and thus the $\chi = 0$ plane represents
a boundary to moduli space, at which the scaling solution
trajectory is reflected. The brane collision occurs as $\psi$, $\chi
\rightarrow - \infty$. Also shown are the directions of increasing
logarithm of the distance between the boundary branes ($\ln{d}$)
and increasing logarithm of the Calabi-Yau volume at the location
of the positive-tension brane ($\phi_+$). Away from the $\chi=0$
boundary, $e^{\phi_+}$ is approximately equal to the volume of the
Calabi-Yau at the location of the negative-tension brane and thus
also approximately
equal to the average Calabi-Yau volume.} }
\end{center}
\end{figure}

\subsection{Recovering the Colliding Brane Solution} \label{sectionScalingSolution}

In \cite{LMT} a colliding branes solution of heterotic M-theory
was derived subject to the boundary conditions that the brane
scale factors and the Calabi-Yau volume should be finite and
non-zero at the collision. This solution was considered in two
different coordinate systems; firstly, one in which the bulk
geometry is static but the branes are moving, and secondly, one in
which the brane locations are fixed and the bulk evolves
dynamically. While in the first coordinate system the solution may
be determined exactly, in the second, comoving, coordinate system,
the solution was found perturbatively as a series expansion in the
rapidity $2 y_0$ of the branes at the collision. The leading term
in this expansion was found to be a scaling solution whose form is
independent of the parameter $y_0$, for any $y_0\ll 1$. It is this
scaling solution that we may expect to recover from the moduli
space description of the system, which holds at low velocities. In
fact, it takes little effort to see that the scaling solution in
\cite{LMT} corresponds to choosing \be
 \t_a = 0, \qquad \psi_0 = 2 y_0, \qquad \chi_0 = 4\a y_0, \qquad
Q_{\psi} = \frac{\sqrt{3}}{2} y_0, \qquad Q_{\chi} = \frac{1}{2}
y_0, \ee from which it follows that \be a= |2 y_0 \t|^{1/2},
\qquad e^{\psi}= |2 y_0 \t|^{\sqrt{3}/4}, \qquad e^{\chi}= |4 \a
y_0 \t|^{1/4}. \label{IntConst2}\ee
As discussed in \cite{LMT}, the condition that the Riemann
curvature of the 5-dimensional bulk remains small, so that
M-theory corrections involving powers of the curvature remain
negligible all the way to the brane collision, selects a unique
solution of the higher-dimensional theory. This solution is also
special in that the brane scale factors and the Calabi-Yau volume
(in 5 dimensions) are finite and non-zero at the collision. In
contrast, from the 4-dimensional perspective, the solution
(\ref{IntConst2}) is just one of an infinite number of seemingly
equivalent solutions, with no special features to distinguish it.
Of course, one could reformulate the requirement of finite
5-dimensional Riemann curvature in terms of 4-dimensional
quantities. But the 5-dimensional interpretation offers more
transparent insight into which solutions do not suffer large
M-theory corrections.

For the solution (\ref{IntConst2}), the brane collision occurs at
time $\t = 0$, where $a=0$ and $\psi$, $\chi \rightarrow -
\infty$. Thus the moduli space scale factor goes to zero at the
collision, whereas we know that from the higher-dimensional point
of view the brane scale factors are finite and non-zero, as may be
verified directly from (\ref{relModScalefactor}).

\begin{figure}[t]
\begin{center}
\includegraphics{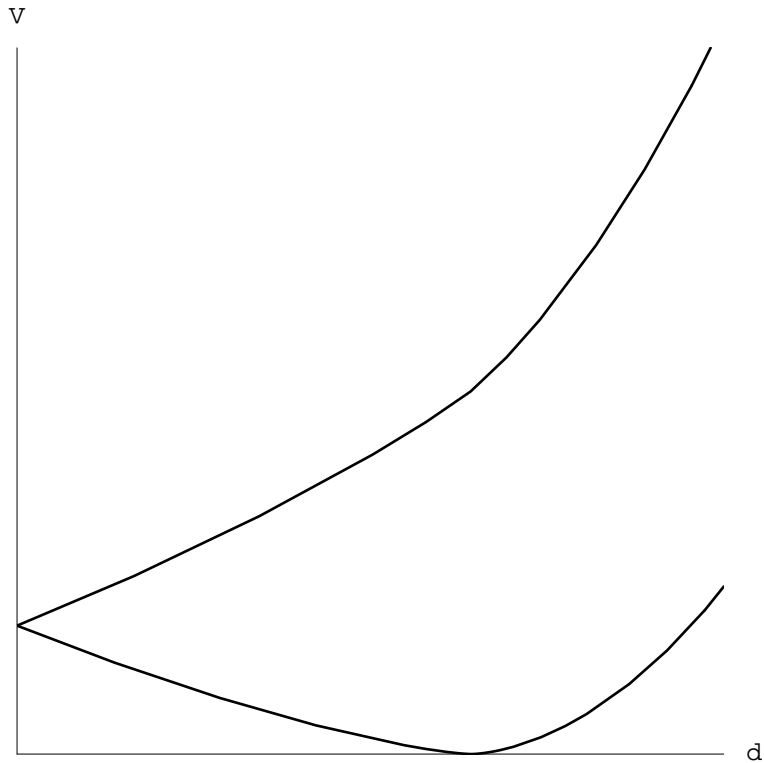} \caption{\label{VdFigure}{\small An alternative viewpoint on
  the colliding branes scaling solution: this plot shows how the volume
$V_{\pm}$ of the Calabi-Yau manifold at the location of the boundary branes at
$y=\pm 1$ depends on the inter-brane distance $d$. The upper curve
  represents $V_+$ while the lower curve represents $V_-.$}}
\end{center}
\end{figure}

The scale factor on the negative-tension brane does however go to
zero at time $\t = 1/(4\a y_0)$, at which time the volume of the
Calabi-Yau manifold also vanishes. This implies the existence of a
boundary to moduli space at $\chi = 0$. As discussed in
\cite{LMT}, we expect the scale factor on the negative-tension
brane to bounce back smoothly when it reaches zero, because of the
peculiar properties of gravity on a brane of negative-tension (in
the presence of matter on the branes, we expect the scale factor
to bounce back {\it before} reaching zero, thus rendering the
bounce entirely non-singular). In comparing the moduli to
higher-dimensional quantities via equations
(\ref{relModDistance})-(\ref{relModScalefactor}), it is apparent
that this bounce of the negative-tension brane is equivalent to
flipping the sign of $\chi$ and thus $\dot{\chi}$ also, and hence
the trajectory of the scaling solution gets reflected off the
$\chi = 0$ boundary.

The scaling solution, as viewed from moduli space, is
shown in Figure \ref{fig1}, where we have suppressed the direction
corresponding to the scale factor $a$. It is immediately apparent
that at the bounce the second derivative of the trajectory is
proportional to a $\delta$-function. Figure \ref{VdFigure} shows
the same solution, but in terms of the physically more meaningful
variables $V_{\mathrm{CY}\pm} = e^{\phi}_{\pm}$, representing the
Calabi-Yau volume at the location of the boundaries at $y=\pm 1$,
and the inter-boundary distance $d$. If $\delta \t$ denotes a
small variation in conformal time about the bounce, then the first
derivative of the $V_-$ curve with respect to $d$ is proportional
to $(\delta \t)^{1/2}$ and is thus zero at the bounce, in
agreement with the fact that the bounce is smooth. However, the
second derivatives with respect to $d$ of both the $V_+$ and the
$V_-$ curves contain a term proportional to $(\delta \t)^{-1/2}$
and thus they blow up at the time of the bounce. This is because
the scaling solution represents only the leading terms of the full
solution expanded in powers of the collision rapidity $2 y_0$. In
the full solution we would expect the trajectories to be rounded
off and to have an everywhere continuous second derivative, for
both choices of variables considered above.

\section{Adding a Bulk Brane} \label{sectionBulk}

From particle phenomenology, we know that generically there are
bulk branes present between the two boundary branes
\cite{Lukas:1998hk}. From the M-theory perspective these bulk
branes arise as M5-branes wrapping a 2-cycle in the Calabi-Yau,
with the remaining four dimensions parallel to the boundaries of
the effective five-dimensional spacetime \cite{Lukas:1998hk}. For
simplicity we will consider adding just a single bulk brane,
although the extension to multiple bulk branes would be
straightforward (though cumbersome) to write down. In the presence
of a bulk brane, the 4-form flux on the Calabi-Yau takes different
values on either side of the bulk brane. The neatest way to handle
this situation is to dualise the flux (which is a scalar in five
dimensions) to a 5-form field strength ${\cal F}=\d{\cal A},$ and
then $\cal{F}$ takes different values on either side of the bulk
brane which is located at $y=Y$ (see \cite{Khoury:2001wf}). The
action reads
\begin{eqnarray}
\nonumber
& & S= \int_{5d} \sqrt{-g}\, [R-\frac{1}{2}(\partial\phi)^2-\frac{3}{2 \cdot 5!}\,e^{2\phi}{\cal F}^2] \nn \\[1ex]
& & \;\;\;\;\;\;- 12 \sum_{i=-,Y,+}\alpha_i \int_{4d, \ y=-1,Y,+1}
[\sqrt{-g} e^{-\phi}
  -\frac{1}{4!}\epsilon^{\mu\nu\kappa\lambda}{\cal A}_{\mu\nu\kappa\lambda}],
\label{eq:5daction}
\end{eqnarray}
where   $\mu,\,\nu,\,\ldots = 0,\, \ldots ,\, 3$. Since the orbifold dimension is
compact, the sum of the tensions $\alpha_i$ must vanish (because
the flux has nowhere to escape). Here we will take
\be
\a_- = -\a_1, \qquad \a_Y = \a_2, \qquad \a_+ = \a_1 - \a_2,
\ee
with $\a_1$ arbitrary and $\a_2$ positive.

\begin{figure}[t]
\begin{center}
\includegraphics{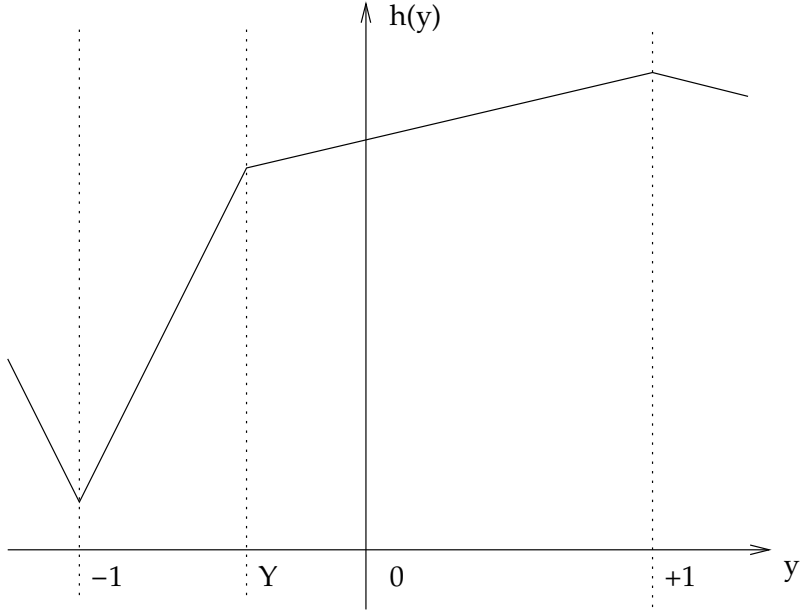} \caption{\label{fig2} {\small The harmonic function in the presence of a bulk brane at $y=Y$.}}
\end{center}
\end{figure}
The static multiple domain wall vacuum solution is then given by
\bea
\d s^2 &=& h^{2/5}(y)\,[B^2(-\d \t^2 + \d \vec{x}^2) + A^2 \d y^2], \\
e^{\f} &=& A\, h^{6/5}(y),  \\[1ex]
 {\cal F}_{0123Y} &=& \begin{cases} & -5\a_1 A^{-1}B^{4}\,h^{-7/5}(y),\;\;\;\;\;\;\;\;\;\;\;\;\;\;\;\;\;\;\; -1 \le y \le Y, \\
& -5(\a_1 -\a_2)\,A^{-1}B^{4}\,h^{-7/5}(y),\;\;\;\;\;\;\;\;\;\; Y \le y \le +1, \end{cases}
\eea
where we have included the (constant) moduli $A$, $B$, $C$ and the new
modulus $Y$. The harmonic function $h$ is now given by
\bea
h&=& \begin{cases} & 5\a_1 y + C, \qquad \qquad \qquad \qquad  \ \  -1 \le y \le Y, \\
& 5(\a_1-\a_2)\,y +C +5\a_2 Y,  \qquad \ Y \le y \le +1,\end{cases}
\eea
where there is now an additional kink at $y=Y$ (see Figure
\ref{fig2}). Proceeding in the same manner as in the last section,
we let the moduli depend on time $\t$, to obtain the moduli space
action
\bea
S_{\mathrm{mod}} &=& -6 \int_{4d}
AB^2I_{\frac{3}{5}}\,\Big[-\frac{1}{12}\Big(\frac{\dot{A}}{A}\Big)^2
+\Big(\frac{\dot{B}}{B}\Big)^2+\frac{\dot{A}\dot{B}}{AB}
-\frac{1}{25}\,\frac{I_{-\frac{7}{5}}}{I_{\frac{3}{5}}}\,\dot{C}^2
+\frac{3}{5}\,\frac{I_{-\frac{2}{5}}\dot{B}\dot{C}}{I_{\frac{3}{5}}B} \nn\\[1ex] && \qquad
+3\a_2\,\frac{I^+_{-\frac{2}{5}}\dot{B}\dot{Y}}{I_{\frac{3}{5}}B}
 -\frac{I^+_{-\frac{7}{5}}}{I_{\frac{3}{5}}}\,\Big(\frac{2\a_2}{5}\,\dot{C}\dot{Y}
+\a_2^2\dot{Y}^2\Big) -\frac{\a_2 \dot{Y}^2}{I_{\frac{3}{5}}(C+5\a_1
  Y)^{\frac{2}{5}}}\Big],
\label{ActionMSA3}
\eea
where the last term originates from the
bulk brane action and where we have introduced the definitions
(note that the definition for $I_n$ is generalised in this
section):
\bea
I^-_{n} &=& \int_{-1}^{Y} dy \ h^n = \frac{1}{5\a_1(n+1)}\,\big[(C+5\a_1 Y)^{(n+1)}-(C-5\a_1)^{(n+1)}\big], \nn \\
I^+_{n} &=& \int_{Y}^{1} dy \ h^n =
\frac{1}{5(\a_1-\a_2)(n+1)}\,\big[(C+5\a_1-5\a_2(1-Y))^{(n+1)}-(C+5\a_1
  Y)^{(n+1)}\big], \nn \\
I_n &=& \int_{-1}^{+1} dy \ h^n = I^-_{n} + I^+_{n}  \\
&=& \frac{1}{5\a_1(\a_1-\a_2)(n+1)} \times \nn \\
&& \big[\a_1(C+5\a_1-5\a_2(1-Y))^{(n+1)}-\a_2(C+5\a_1
Y)^{(n+1)}-(\a_1-\a_2)(C-5\a_1)^{(n+1)}\big]. \nn \eea Once again we
can define an effective four-dimensional scale factor $a$ via \be a^2
= AB^2 I_{\frac{3}{5}}, \ee and use the expression \be \dot{I}_n =
n \,\dot{C} I_{(n-1)} +  5\a_2 n\,  \dot{Y} I^+_{(n-1)} \ee in
order to rewrite the moduli space action as \bea S_{\mathrm{mod}}
&=& \int_{4d} -6 \dot{a}^2 + a^2
\Big[2\Big(\frac{\dot{A}}{A}\Big)^2 + \frac{27}{50}\,
\Big(\frac{I_{-\frac{2}{5}}}{I_{\frac{3}{5}}}\Big)^2 \dot{C}^2
+\frac{6}{25}\,\frac{I_{-\frac{7}{5}}}{I_{\frac{3}{5}}}\,\dot{C}^2 \nn \\[1ex]
&& + \frac{6 \a_2}{I_{\frac{3}{5}}(C+5\a_1
Y)^{\frac{2}{5}}}\,\dot{Y}^2 + 6 \a_2^2\,
\frac{I^+_{-\frac{7}{5}}}{I_{\frac{3}{5}}}\,\dot{Y}^2
+\frac{27\a_2^2}{2}\,
\Big(\frac{I^+_{-\frac{2}{5}}}{I_{\frac{3}{5}}}\Big)^2\,\dot{Y}^2 \nn \\[1ex] &&
+\frac{9}{5}\,
\frac{I_{-\frac{2}{5}}}{I_{\frac{3}{5}}}\,\frac{\dot{A}}{A}\,\dot{C}
+9\a_2\,
\frac{I^+_{-\frac{2}{5}}}{I_{\frac{3}{5}}}\,\frac{\dot{A}}{A}\,\dot{Y}
+\frac{12\a_2}{5}\,
\frac{I^+_{-\frac{7}{5}}}{I_{\frac{3}{5}}}\,\dot{C}\dot{Y}
+\frac{27\a_2}{5} \,\frac{I_{-\frac{2}{5}}
I^+_{-\frac{2}{5}}}{(I_{\frac{3}{5}})^2}\,\dot{C}\dot{Y}\Big].\qquad
\label{ActionMSA4} \eea This action describes gravity with scale
factor $a$ coupled to three scalar fields $A$, $C$ and $Y$. In
analogy with the case where no bulk brane is present, one would
hope to be able to reduce the action to a much simpler form by a
series of field redefinitions. However, in the present case it
seems unlikely that such a drastic simplification can be achieved,
since a calculation of the curvature of the scalar field manifold
inhabited by $A$, $C$ and $Y$ reveals the Ricci scalar to be a rather
complicated function of $C$ and $Y$ (by contrast, in the absence
of a bulk brane, the scalar field manifold is flat). In view of
this difficulty, we will simplify the action by looking at certain
specific limits in the following subsections.

Before doing so, however, we would like to remark that the moduli
space under consideration here has two very obvious boundaries in
the $Y$ direction, namely at $Y = -1$ and at $Y = +1,$
corresponding to the collision of the bulk brane with the boundary
branes. It is not clear, however, to what extent the moduli space
action can be trusted at these specific boundaries, since a bulk
brane could fuse momentarily with the boundary, accompanied by a
small instanton transition \cite{Ovrut:2000qi}. This specific
process would then not be described by the moduli space
approximation, as one would expect that additional light degrees
of freedom, describing the interaction of the bulk brane with the
boundary brane, would have to be added to the effective action.

\subsection{The Large Harmonic Function Limit}

For cosmological applications, and in particular applications to
the ekpyrotic/cyclic models, the most useful regime to consider is
the one where the boundary branes are far apart and slowly
approaching one another. Indeed, this corresponds to the epoch
where the cosmological density perturbations are being generated.
From the colliding brane scaling solution described in Section
\ref{sectionScalingSolution}, in conjunction with
(\ref{relModDistance}) and (\ref{relCmoduli}), it is easy to see
that the large $|\t|$ limit corresponds to the large inter-brane
distance limit, which in turn corresponds to the modulus $C(\t)$
being very large. In fact, $C(\t)$ has the property that it also
becomes very large in the near-collision limit $\t \rightarrow 0.$
Thus both the near-collision limit and the large boundary
separation limit correspond to the harmonic function $h$ being
very large. We can expand the relevant integrals in powers of $C$:
\bea
I^-_{n} &=& C^n (1+Y) + \frac{5n\a_1}{2}\,C^{(n-1)}(Y^2-1) + {\cal{O}}(C^{(n-2)}), \\
I^+_{n} &=& C^n (1-Y) +\frac{5n}{2}\,C^{(n-1)}[(\a_1+\a_2)Y+\a_1-\a_2] + {\cal{O}}(C^{(n-2)}),\\
I_{n} &=& 2 C^n -\frac{5n\a_2}{2} \,C^{(n-1)}(1-Y)^2 + {\cal{O}}(C^{(n-2)}).
\eea
We then expand the moduli space action in powers of $1/C$,
and to first order we find
 \bea
S_{\mathrm{mod}, \,C \rightarrow \infty} &=&  \int_{4d}
-6 \dot{a}^2 + a^2 \Big[2\Big(\frac{\dot{A}}{A}\Big)^2 +
\Big(\frac{39}{50}+\frac{39\a_2}{20 C}\,(1-Y)^2 \Big)
\Big(\frac{\dot{C}}{C}\Big)^2 + \frac{3\a_2}{C}\,\dot{Y}^2 \nn \\[1ex] &&
+\Big(\frac{9}{5}+\frac{9\a_2}{4C}\,(1-Y)^2\Big)\,\frac{\dot{A}\dot{C}}{AC}
+\frac{9\a_2(1-Y)}{2} \,\frac{\dot{A}\dot{Y}}{AC}
+\frac{39\a_2(1-Y)}{10}\, \frac{\dot{C}\dot{Y}}{C^2}\Big].\qquad
\label{ActionMSA5}
\eea
In order to get a better physical
understanding of the meaning of the various terms in this action,
we will define the two fields
\bea
D &\equiv& \ln (AC^{1/5}), \\
\f_A &\equiv& \ln (A C^{6/5}).
\eea
These fields are related to the distance between the boundary branes,
and the orbifold-averaged Calabi-Yau volume,
in the limit of $C$ being large:
\bea
d &\simeq& 2 e^D, \\
\langle V\rangle &\simeq& e^{\f_A}.
\eea
Note also that in the limit we
are considering
\be
\f_A \gg D \gg 1.
\ee
In terms of these new fields, the moduli space action reads
\bea
S_{\mathrm{mod},\, C \rightarrow
\infty} &=& \int_{4d} -6 \dot{a}^2 + a^2 \Big[\frac{3}{2}\,\dot{D}^2 +
\frac{1}{2}\, \dot{\f}_A^2 + 3\a_2 \,e^{D-\f_A}\, \dot{Y}^2 +
\frac{3\a_2}{2}\,(1-Y)\,e^{D-\f_A}\,(\dot{D}+2\dot{\f}_A) \,\dot{Y}
 \nn \\ &&\qquad + \frac{3\a_2}{4}\,(1-Y)^2 \,e^{D-\f_A}\,(2\dot{\f}_A^2-\dot{\f}_A\dot{D}-\dot{D}^2)\Big]
  + {\cal{O}}(e^{2(D-\f_A)}).
\label{ActionMSA6} \eea Note that if the bulk brane is very close
to the boundary at $y=1$ and we momentarily insert $Y=1$ into the
action, we obtain \bea S_{\mathrm{mod}, \, C \rightarrow \infty,\,
Y=1} &=& \int_{4d} -6 \dot{a}^2 + a^2 \Big[\frac{3}{2}\,\dot{D}^2
+ \frac{1}{2}\, {\dot{\f}_A}^2 + \frac{3\a_2}{5}\, e^{D-\f_A}\,
\dot{Y}^2 \Big] , \label{ActionMSA7} \eea which is the action
previously derived (by different means) in
\cite{Derendinger:2000gy,Brandle:2001ts}. However, inserting $Y=1$
into the action is of course inconsistent\footnote{In this
particular case it turns out that the Ricci scalar on the scalar
field manifold calculated from the action (\ref{ActionMSA7})
coincides with the Ricci scalar calculated using the full action
(\ref{ActionMSA6}) and inserting $Y=1$ at the end (in both cases
its value is $-4/3$). However, this is a coincidence; the
connections, for example, are not equal.}: one should use the full
action (\ref{ActionMSA6}) and only at the end of a calculation
insert particular values of $Y$. However, as mentioned previously,
the moduli space approximation is unlikely to be very trustworthy
at this particular boundary in any case.

\subsection{The Symmetric Case}

As another example where the moduli space action simplifies
considerably, we will consider the symmetric case with two
boundary branes both with negative tension $-\a_1$ and one bulk
brane with positive tension $\a_2 = 2 \a_1,$ the bulk brane being
located near $Y=0.$ We are trying to find the effective action for
this setup in the limit where the boundaries are far apart, but
only slowly moving, and we will also specialise to the
phenomenologically interesting limit where the orbifold-averaged
Calabi-Yau volume is fixed.

Writing $h_{\pm} \equiv h(y=\pm 1)$ and $h_Y \equiv h(y=Y)$,
we have
\bea
I^+_{n} &=& \frac{1}{5\a_1 (n+1)}\, [h_Y^{(n+1)} - h_+^{(n+1)}], \\
I_{n} &=& \frac{1}{5\a_1 (n+1)}\, [2 h_Y^{(n+1)} - h_+^{(n+1)} - h_-^{(n+1)}].
\eea
The large boundary separation limit is obtained
by letting $h_{\pm} \rightarrow 0$.  Taking this limit corresponds
to $C \rightarrow 5\a_1$ and thus $h_Y = h_0 \simeq 5\a_1$. We
then obtain the following expressions for various integrals of the
harmonic function:
\bea
  n>-1:  \qquad I^+_{n} &\simeq&
\frac{1}{5\a_1 (n+1)}\, h_Y^{(n+1)},
\\ n>-1: \qquad I_{n}
&\simeq& \frac{2}{5\a_1 (n+1)} \,h_Y^{(n+1)},
 \\n<-1: \qquad I^+_{n}
&\simeq& \frac{-1}{5\a_1 (n+1)}\, h_+^{(n+1)},
\\n<-1: \qquad I_{n}
&\simeq& \frac{-1}{5\a_1 (n+1)} \,[h_+^{(n+1)} + h_-^{(n+1)}],
\eea
while the orbifold-averaged Calabi-Yau volume reduces to
\be
\langle V_{CY}\rangle = \frac{10}{11}\, A\, (5\a_1)^{6/5}.
\ee
This suggests we should take $A$ to be constant, so that the average
Calabi-Yau volume is fixed. If we retain only the leading terms,
the moduli space action becomes
\bea
S_{\mathrm{mod, \,\,symmetric}} &=&
\int_{4d} -6\dot{a}^2 + a^2\, \Big[\frac{12}{25 \, h_0^{8/5}} \, \Big(
\frac{\dot{h}_-^2}{h_-^{2/5}} +
\frac{\dot{h}_+^2}{h_+^{2/5}}\Big)  \Big].
\label{ActionMSA_sym1}
\eea
From an eleven-dimensional point of view
the distance between the boundary brane at $y = -1$ and the bulk
brane at $y =0$ is given by ({\it cf.} (\ref{distance11d}))
\be
d_{11,\,-0} = \frac{1}{4\a_1} \,A^{2/3} \, (h_0^{4/5} - h_-^{4/5}).
\ee
In the limit under consideration, we obtain the approximate relationship
\be
\frac{\dot{d}_{11,\,-0}}{d_{11,\,-0}} \simeq - \frac{4}{5}\,
\frac{\dot{h}_-}{h_0^{4/5} h_-^{1/5}},
\ee
and similarly for $d_{11,\,+0}$. Thus the moduli space action can be
rewritten as
\bea
S_{\mathrm{mod,\,\, symmetric}} &=&  \int_{4d} -6\dot{a}^2 +
a^2 \,\Big[\frac{3}{4} \, \Big( \frac{\dot{d}_{11,\,-0}^2}{d_{11,\,-0}^2} +
\frac{\dot{d}_{11,\,+0}^2}{d_{11,\,+0}^2} \Big)  \Big],
\label{ActionMSA_sym2}
\eea
describing gravity minimally coupled to two scalar fields
representing the distance of the bulk brane to the two respective
boundaries. As expected, the moduli space action embodies the
symmetries of the specific setup analysed here.

\section{Conclusions} \label{sectionDiscussion}

We have developed the moduli space approximation for heterotic
M-theory, including the case where a bulk brane is moving along
the orbifold direction. The moduli space actions describe gravity
in the form of an effective scale factor coupled to scalar fields.
In general the resulting equations of motion allow for a very
large number of possible motions of the branes. In this context,
the boundary conditions that one obtains by inspection of the
higher-dimensional parent theory can prove crucial in singling out
a particularly relevant solution. Moreover, the parent theory is
useful in determining what the allowed ranges of the moduli are;
we have given examples of such boundaries to moduli space, and
shown how these result in important modifications to the allowed
moduli space trajectories.

In the absence of a bulk brane, the action is remarkably simple,
and consists of gravity minimally coupled to two scalar fields
(which only interact with each other via gravitational effects).
At large brane separation the respective logarithms of the
distance between the branes and the Calabi-Yau volume are
orthogonal variables, and one can perform a rotation in field
space to obtain \bea S_{\mathrm{mod},\, C \rightarrow \infty} &=& \int_{4d}
-6 \dot{a}^2 + a^2 \,\Big[\frac{3}{2}\,\dot{D}^2 + \frac{1}{2}\,
{\dot{\f}_A}^2\Big], \eea where the distance between the branes is $2\,
e^D$ and the average Calabi-Yau volume is given by $ e^{\f_A}.$
One of the allowed trajectories in moduli space corresponds to the
colliding brane solution described in \cite{LMT}. This solution
has been proposed as a model for the big bang, because it has the
particular feature that the brane collision is well-behaved, in
the sense that
the Riemann curvature remains bounded right up to (but not
including) the collision. It would be of interest to study the
generation of cosmological perturbations in this model. However, in
order to do so one would have to know the inter-brane forces away
from the collision. In complicated settings such as heterotic
M-theory, where the forces between the various branes can only
partly be computed as yet (see {\it e.g.}
\cite{Lima:2001jc,Moore:2000fs}), the moduli space approximation
offers us the possibility of adding ``effective'' potentials for
the moduli to the effective action. These effective potentials
then reflect our guess for the sum of all inter-brane forces and
allow one to deduce the resulting spectrum of cosmological
perturbations. Work related to these questions will be presented
elsewhere \cite{LMTS}.

In the presence of a bulk brane, the moduli space action is in general quite
complicated, although we have shown how it simplifies in certain limits
of interest.
In this way we have
been able to compare our work with results already known in the
literature, and to extend these. The appearance of an extra scalar
degree of freedom is likely to have interesting phenomenological
consequences.

\begin{center}
***
\end{center}
{\it Acknowledgements:}
The authors wish to thank Andre Lukas, Ian Moss and Andrew Tolley
for useful discussions. JLL and NT acknowledge the support of
PPARC and of the Centre for Theoretical Cosmology, in Cambridge.
PLM is supported through a Spinoza Grant of the Dutch Science
Organisation (NWO).

\bibliographystyle{apsrev}
\bibliography{DraftActions}

\end{document}